\documentclass[apj]{emulateapj}

\newcommand{\geqsim}{\,\raisebox{-0.6ex}{$\buildrel > \over \sim$}\,}

\shorttitle{$Spitzer$ Study of MS1512-cB58}
\shortauthors{Siana et al.}

\begin{document}


\title{Spitzer Observations of the $z=2.73$ Lensed Lyman Break Galaxy, MS1512-cB58}


\author{Brian Siana\altaffilmark{1}, Harry I. Teplitz\altaffilmark{2}, Ranga-Ram Chary\altaffilmark{2}, James W. Colbert\altaffilmark{2}, David T. Frayer\altaffilmark{3}}
\altaffiltext{1}{California Institute of Technology, MS 105-24, Pasadena, CA 91125}
\altaffiltext{2}{$Spitzer$ Science Center, California Institute of Technology, MS 220-6, Pasadena, CA 91125}
\altaffiltext{3}{Infrared Processing and Analysis Center, California Institute of Technology, MS 100-22, Pasadena, CA 91125}
\email{bsiana@astro.caltech.edu}

\begin{abstract}
We present Spitzer infrared (IR) photometry and spectroscopy of the lensed Lyman break galaxy (LBG), MS1512-cB58 at $z=2.73$.  The large (factor $\sim 30$) magnification allows for the most detailed infrared study of an $L^*_{UV}(z=3)$ LBG to date.  Broadband photometry with IRAC (3-10 $\mu$m), IRS (16 $\mu$m), and MIPS (24, 70 \& 160 $\mu$m) was obtained as well as IRS spectroscopy spanning 5.5-35 $\mu$m.    A fit of stellar population models to the optical/near-IR/IRAC photometry gives a young age ($\sim9$ Myr), forrming stars at $\sim 98$ M$_{\odot}$ yr$^{-1}$, with a total stellar mass of $\sim 10^9$ M$_{\odot}$ formed thus far.  The existence of an old stellar population with twice the stellar mass can not be ruled out.  IR spectral energy distribution fits to the 24 and 70 $\mu$m photometry, as well as previously obtained submm/mm, data give an intrinsic IR luminosity $L_{IR} = 1-2 \times 10^{11}$ L$_{\odot}$ and a star formation rate, SFR $\sim20-40$ M$_{\odot}$ yr$^{-1}$.  The UV derived star formation rate (SFR) is $\sim3-5$ times higher than the SFR determined using $L_{IR}$ or $L_{H\alpha}$ because the red UV spectral slope is significantly over predicting the level of dust extinction.  This suggests that the assumed Calzetti starburst obscuration law may not be valid for young LBGs.  We detect strong line emission from Polycyclic Aromatic Hydrocarbons (PAHs) at 6.2, 7.7, and 8.6 $\mu$m.  The line ratios are consistent with ratios observed in both local and high redshift starbursts.  Both the PAH and rest-frame 8 $\mu$m luminosities predict the total $L_{IR}$ based on previously measured relations in starbursts.  Finally, we do not detect the 3.3 $\mu$m PAH feature.  This is marginally inconsistent with some PAH emission models, but still consistent with PAH ratios measured in many local star-forming galaxies.  

\end{abstract}

\keywords{galaxies: high-redshift, galaxies: individual (MS1512-cB58), infrared: galaxies}

\section{Introduction}
Much of the global star formation at $z>2$ occurs in ultraviolet-luminous Lyman Break Galaxies \citep[LBGs,][]{steidel96,reddy05}.  For most LBGs, much of what we know about their star-formation (star formation rates, dust reddening) is derived from rest-frame UV properties, where considerable degeneracies exist (ie. dust reddening and starburst age).  In most $L^*_{UV}$ LBGs, the majority of the UV photons are absorbed by dust \citep{adelberger00,reddy08}, which then emits the energy in the infrared.  Therefore, an accurate measurement of the total infrared luminosities would give a complete census of the reprocessed UV photons and thus, a better determination of the star formation rates in LBGs.  LBGs can be detected in the infrared at 24 $\mu$m with the {\it Spitzer Space Telescope}.  Observations in the $Spitzer$ 24 $\mu$m bandpass are the most sensitive towards detecting dust obscured star formation at $z<3$.  However, even the deepest $Spitzer$ surveys ($f_{24}>20$ $\mu$Jy, or $L_{IR}\geqsim3\times10^{11}$ L$_{\odot}$ at $z=3$), can not detect the majority of LBGs \citep[see IR luminosity function of][]{reddy08}.  Between $1<z<3$, the 24 $\mu$m flux is dominated by the emission features of Polycyclic Aromatic Hydrocarbons (PAHs), resulting in uncertain bolometric corrections due to variations in the PAH emission and possible silicate absorption at 9.7 $\mu$m.  As the IR spectra of typical LBGs have not been measured, templates of local starbursts have been used to extrapolate the rest-frame 5-12 $\mu$m luminosities to total IR luminosities.  It is not clear, however, that the IR SEDs, and in particular the PAH emission, is the same in local and high redshift starbursts.  

Much progress has been made in measuring the mid-IR spectral properties and broadband SEDs of dusty, ultraluminous star-forming galaxies at $z>1$ \citep{yan05,menendez-delmestre07,sajina07,desai07,pope08}, and even an extremely luminous and dusty LBG \citep{huang07}.  However, there have been no mid-IR spectra obtained of LBGs with typical luminosities ($M_{1500} \sim -21$, $L_{IR}\sim10^{11}$ L$_{\odot}$) and dust extinction ($0.0<E(B-V)<0.3$).  There are a few LBGs that are sufficiently magnified through lensing for their mid-IR spectral properties and far-IR photometry to be measured \citep[e.g.,][]{yee96,smail07,allam07}.  MS1512-cB58 (hereafter, cB58) is the first lensed LBG \citep[$z=2.73$,][]{yee96} to be found with a large magnification \citep[$\sim30$,][]{williams96_cb58,seitz98}, allowing extensive follow-up at various wavelengths.  Fits of stellar population models to the optical and near-IR photometry (rest-frame UV and optical) indicate a very young ($t_{age} \sim 10-20$ Myr), relatively dusty, $E(B-V) \sim 0.3$, starburst \citep{ellingson96}.  \citet{pettini00} use the rest-frame UV continuum to determine the dust extinction and derive a reddening (and magnification) corrected star formation rate of 93 $M_{\odot}$ yr$^{-1}$.  A Keck near-IR spectrum of nebular emission lines yields an intrinsic star formation rate of $\sim 20$ M$_{\odot}$ yr$^{-1}$ \citep{teplitz00}.  Sub-millimeter \citep{van_der_werf01,sawicki01} and millimeter \citep{baker01} photometry also suggest, though with large uncertainty, a star formation rate significantly lower than the UV-derived SFR, calling into question the applicability of locally derived relations (reddening laws, IR SEDs) to this LBG, and perhaps the population as a whole.  However, this discrepancy can be reconciled if the IR SED is very warm, as predicted by the LBG dust emission models of \citet{takeuchi04}.  Here we present Spitzer mid- and far-IR photometry, and mid-IR spectroscopy of cB58 to more accurately determine its IR properties and whether local starburst relations can be used for analysis of LBGs.  

We use a $\Lambda$CDM cosmology with $\Omega_m = 0.3$, $\Omega_{\Lambda}=0.7$, and $H_0 = 70$ km s$^{-1}$ Mpc$^{-1}$.  All intrinsic luminosities and star-formation rates are corrected assuming a lensing magnification, $\mu = 30$ \citep{seitz98}.  \citet{seitz98} estimate an error on the magnification 30\%.

\section{Observations and Data Reduction}
\label{obs_reduc}

Both IRAC and MIPS observations were part of a GTO program (PID:65, PI: C. Lawrence).  cB58 was imaged in all four bands of the IRAC instrument \citep[3.6/4.5/5.8/8.0 $\mu$m, ][]{fazio04} 2004 January 18.  In each band, the target was imaged at nine dither positions for a total of 900s per band.  Far-IR photometry with MIPS \citep{rieke04} was taken 2004 March 20.  The 24, 70, and 160 $\mu$m bands were imaged for 2263, 1091, and 105 seconds, respectively.  

\begin{figure*}
\epsscale{1.1}
\plotone{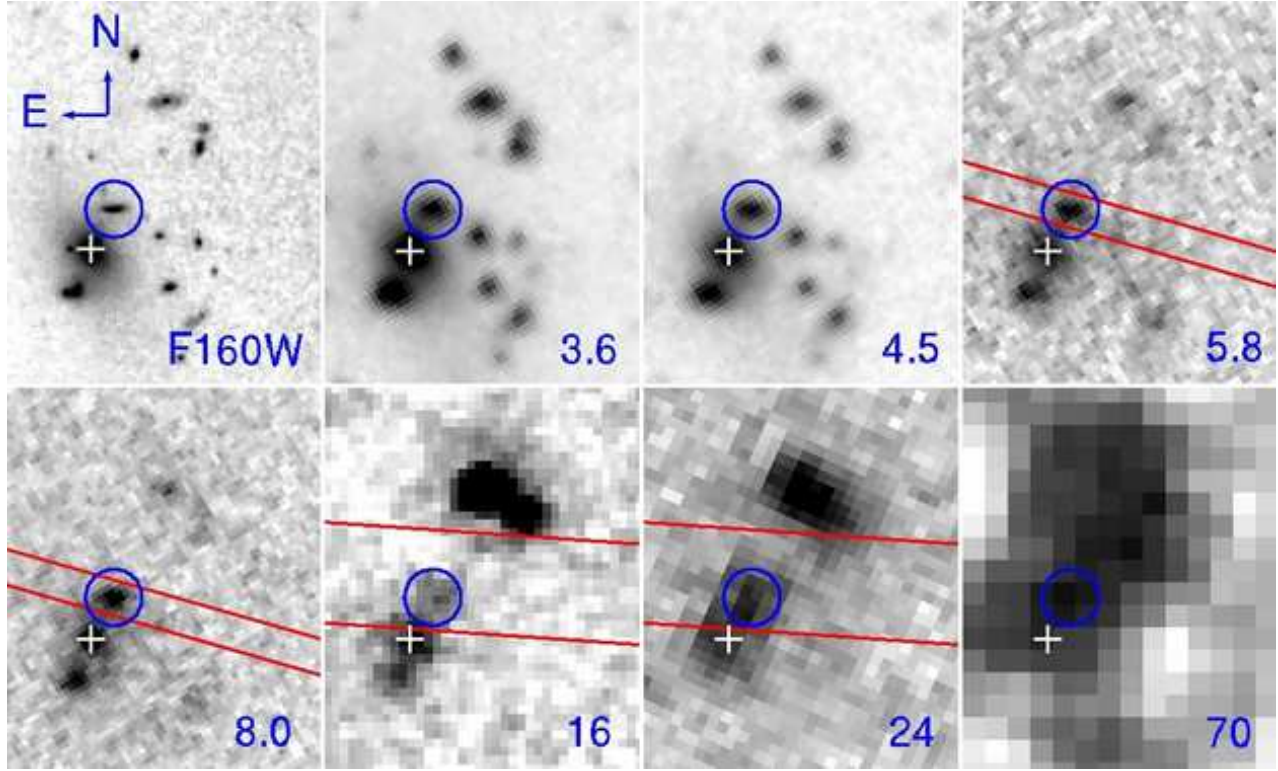}
\caption{HST/IRAC/MIPS postage stamps of MS1512-cB58.  From top left: HST NICMOS F160W (H-band) followed by the four IRAC bands, IRS 16 $\mu$m, MIPS 24 and 70 $\mu$m.  cB58 is marked with a circle of 3$''$ radius, and the cluster cD galaxy is marked with a white cross.  The cluster cD and cB58 are $\sim 5.3''$ apart.  The slit positions and position angles are shown in the images at the appropriate wavelengths (Short-Low in IRAC 5.8 \& 8.0 $\mu$m, and Long-Low in IRS 16 and MIPS 24).  The Long-Low slit is offset 2.35$''$ from cB58 in the direction opposite from the cD galaxy (i.e. 27.5$^{\circ}$ W of N).  The postage stamps are 34$''$x41$''$ and North is up, East is left. \label{fig:stamps}}
\end{figure*}

All IRS observations were taken for Spitzer Program 30832.  We obtained 16 $\mu$m with the Blue Peak-Up imager on the $Spitzer$ IRS instrument on 2007 June 25 for a total exposure time of 1415 seconds.  IRS spectral observations were performed on 2007 March 23 (Long-Low) and 2007 June 21 (Short-Low).  The spectroscopy was taken in mapping mode, placing the target in four different positions along the Short-Low slit and six positions along the Long-Low slit as recommended by \citet{teplitz07} for optimal $S/N$ in deep IRS integrations.   All IRS observations were scheduled immediately after the ``skydark'' calibrations to reduce the chance of latent images from preceding bright targets.  Total integration time was 6.77 ks for both SL1 and SL2, 14.63 ks in LL2, and 34.13 ks in LL1.  

cB58 is only 5.3$''$ away from the cD galaxy of the foreground cluster ($z=0.373$) and this galaxy has approximately the same flux at 24 $\mu$m (250 $\mu$Jy) as cB58, so extra caution was required to avoid light from the cD galaxy falling in the IRS slits.  For both Short-Low and Long-Low observations, the slit was aligned as close to perpendicular with the cB58-cD connector as was allowed given scheduling constraints (see Figure \ref{fig:stamps}).  Since the Long-Low slit is 10.7$''$ wide (approximately twice the distance between cB58 and the cD galaxy), half of the light from the cD galaxy would have fallen into the slit as well.  Therefore, we offset the IRS Long-Low observations 2.35$''$ away from the cD galaxy.  Given the point-spread function, we estimate that this eliminates $\sim87$\% of the cD galaxy light while only decreasing the cB58 flux by $\sim6$\%.  

The IRAC data were reduced using the SSC's MOPEX software and drizzled on to a finer pixel scale (0.6$''$ pix$^{-1}$).  The cD galaxy is bright in the IRAC images and it's profile clearly extends beyond cB58 (Figure \ref{fig:stamps}).  We use the GALFIT package \citep{peng02} to fit a de Vaucouleurs profile to the cD galaxy and subtract it from the images.  We then extract the flux in a large elliptical (8$''$x6$''$) aperture (cB58 is marginally extended).  We do not apply aperture corrections because these are small ($<3$\%).  At 3.6 and 4.5 $\mu$m, where the cD galaxy is brightest, GALFIT has slightly oversubtracted the profile of the cD galaxy in the region of cB58.  Therefore the background is slightly lower around cB58.  We add a small constant to the background to correct for this, resulting in a $\sim5$\% increase in the cB58 fluxes at 3.6 \& 4.5 $\mu$m.  We added a 5\% systematic error (to account for the uncertain background estimation) in quadrature with the statistical errors and the uncertainty in the absolute IRAC calibration (5\%).
 
The IRS 16 $\mu$m and MIPS 24 $\mu$m data were reduced with the MOPEX package.  The 70 $\mu$m data were reduced using the Germanium Reprocessing Tools (GeRT), following the techniques optimized for deep photometry data given by \citet{frayer06}.  The IRS 16, and MIPS 24 \& 70 $\mu$m images are shown in Figure \ref{fig:stamps}.  The PSF at  16 $\mu$m is sufficiently small to resolve cB58 from the cD galaxy.  In both the 24 and 70 $\mu$m images, confusion with nearby sources is a concern.  At 24 $\mu$m, cB58 is marginally confused with the cD galaxy, 5.3$''$ to the SE (the 24 $\mu$m FWHM is 6$''$).  We use the IRAC positions as priors to simultaneously fit two point sources at these positions, giving $f_{24}(cB58) = 240 \pm 40$ $\mu$Jy.  At 70 $\mu$m, the confusion is more severe, as there is also a bright 24/70 micron source $\sim12''$ N-NW of cB58 (the 70 $\mu$m FWHM is 18$''$).  Three point sources are fit simultaneously at the 24 $\mu$m positions, giving a $f_{70}(cB58) = 1.7 \pm 1.0$ mJy. The cD and the northern galaxy are both brighter (2.7 and 4.5 mJy, respectively).  The error includes the error in the multiple source fit.  If we do only a two component fit with the bright NW galaxy and the combined cB58+cD, we get a $f_{70}(cB58+cD) = 4.7$ mJy, a conservative upper limit to the cB58 flux.  We do not detect cB58 at 160 $\mu$m and derive a $3\sigma$ upper limit of $f_{160} < 24$ mJy based on the confusion limit of \citet{dole04}.

IRS data reduction was performed as specified in \citet{teplitz07}.  First, we removed latent charge by fitting the slope of the increase in backround with time and subtracting this background row by row.  Second, ``rogue'' pixels were masked using the IRSCLEAN program provided by the SSC.  Finally, the observations at other map positions were used to determine the sky, which was then subtracted.  The individual frames were co-added to produce 2D spectra at each map position.  One-dimensional spectra were optimally extracted at each map position using the SPICE software provided by the SSC.  

As mentioned above, special consideration is required to account for contamination of the cB58 spectrum by the cluster cD galaxy.  Contamination is primarily of concern at longer wavelengths (Long-Low spectrum) because of the broader PSF, and the wider slit width (10.7$''$).  We estimate the contribution of the cD galaxy to the IRS Long-Low spectrum in the following manner.  First, we use the SSC's model of the 24 $\mu$m PSF (for a $T=50$K blackbody spectrum) and determine the fraction of the flux falling into the Long-Low slit for several slit positions.  For a slit offset from the center of the PSF by 2.35$''$ (the Long-Low position offset from cB58), we detect $\sim93.6$\% of the flux that is detected when the source is centered within the slit.  For a slit offset 7.35$''$ from the PSF center (the total offset from the central cD galaxy), we detect only $\sim23.4$\% of the flux detected when the source is centered within the slit.  Because the 24 $\mu$m fluxes of the two sources are known ($f_{24}(cB58)=240$ \& $f_{24}(cD)=250$ $\mu$Jy), we determine that the resulting spectrum will have a contribution of $250\times0.234 = 59$ $\mu$Jy from the cD galaxy and $240\times0.936=222 $ $\mu$Jy from cB58 for a total of 281 $\mu$Jy.  This agrees very well with the extracted spectrum flux (uncorrected for contamination) $f_{24}=284$ $\mu$Jy, when multiplying by the MIPS 24 $\mu$m transmission curve and integrating over the bandpass.  Therefore, to correct the spectrum for contamination from the cD galaxy, we subtract 23.4\% of the estimated cD galaxy spectrum.  Then, to correct for slit loss of the cB58 flux we multiply the spectrum by 1.064.   

\begin{equation}
f_{\nu}(cB58) = 1.064[f_{\nu}(obs)-0.234f_{\nu}(cD)]
\end{equation}

The cD galaxy spectrum, $f_{\nu}$(cD), is assumed to be a power law fit to the 16 \& 24 $\mu$m fluxes because, at the redshift of the cD galaxy ($z=0.37$), the IRS Long-Low 1st order spectrum probes wavelengths ($16<\lambda_{rest}<24$ $\mu$m) where strong emission lines or absorption features are not typically seen in either star-forming galaxies or AGN \citep{armus04, brandl06, buchanan06,netzer07}.  The net effect is that the continuum is supressed by 10-20\% while the PAH fluxes remain largely unchanged.

This exercise only tells us the fractional contamination levels at $\sim 24$ $\mu$m, but the contamination is a function of wavelength because the PSF width is a function of wavelength. The contamination fraction was computed over the large 24 $\mu$m bandpass (21-27 $\mu$Jy) so the PSF will not be significantly different between the 20-35 $\mu$m region of interest.  Therefore we do {\it not} estimate a wavelength dependent contamination correction as this is a second order correction. 

\section{Results}

\subsection{Stellar Population Fits}

\begin{figure}
\epsscale{1.0}
\plotone{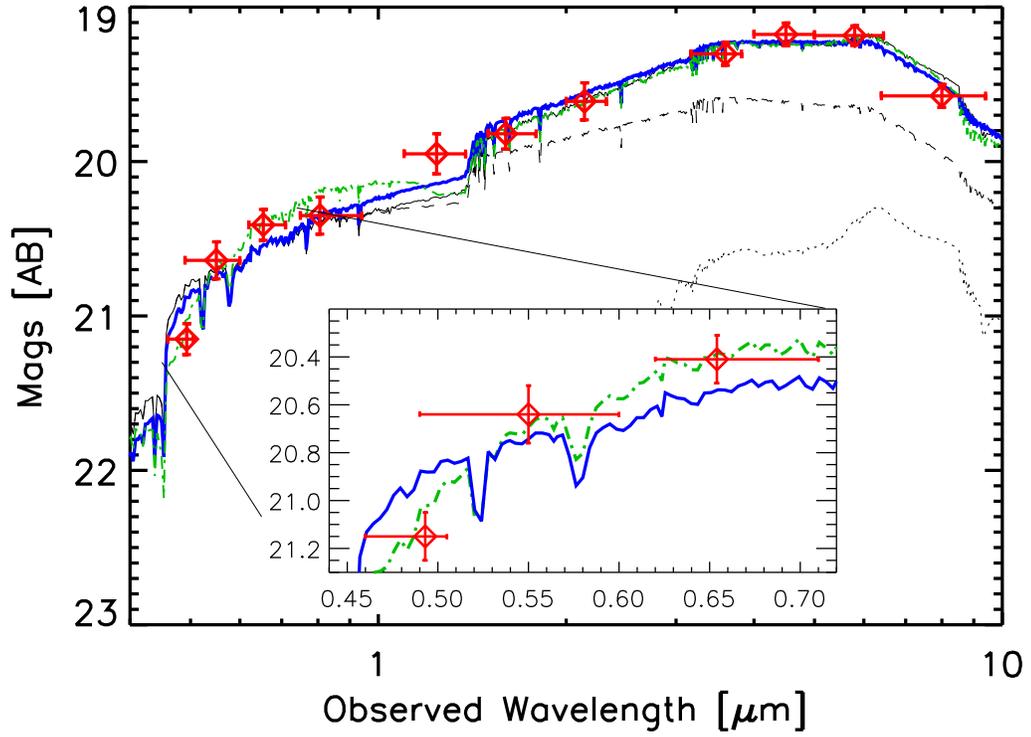}
\caption{The SED fit to the optical/near-IR/mid-IR photometry of cB58.  The solid thick blue line is the single component fit using a Calzetti obscuration law.  The thin solid line is the sum of two components: a continuous star formation component (dashed) and an old (2 Gyr) single burst component (dotted).  The combined two component SED is similar to the single component SED and is therefore hidden under the thick blue solid line.  Also plotted (green dot-dashed line) is the single component fit with an LMC extinction curve (without the 2175 \AA\ feature) that better produces the red UV spectral slope (shown in the inset).  \label{fig:sedfit}}
\end{figure}

The $Spitzer$ photometry of cB58 is given in Table \ref{tab:photom}.  The IRAC photometry traces the rest-frame near-IR, allowing us to more accurately determine the stellar population parameters than with rest-frame UV/optical data alone.  We combine the optical and near-IR photometry from \citet{yee96} and \citet{ellingson96}, respectively, with the IRAC photometry to fit the complete stellar SED from rest-frame 1200 \AA\ to 2 $\mu$m.  We use population synthesis models from \citet{bruzual07}, which account for the thermally pulsing AGB phase of stellar evolution.  Three different star formation histories (constant, single burst, exponential decay with $e$-folding time, $\tau = 100$ Myr) are used and the age of the burst is allowed to vary. The dust obscuration law for starburst galaxies \citep{calzetti97}, a Salpter IMF \citep{salpeter55} and a metallicity $Z=0.4 Z_{\odot}$ is assumed, similar to previously determined cB58 metallicities \citep{pettini00,teplitz00}.  We find that a very young, constant star formation model is required to explain the lack of a Balmer break between the J and H bands.  The best-fit SED, shown in Figure \ref{fig:sedfit}, has an age of $9.3^{+4.7}_{-3.1}$ Myr, $SFR = 98^{+46}_{-31}$ M$_{\odot}$ yr$^{-1}$, an  above average reddening for an LBG, $E(B-V) = 0.35 \pm 0.03$, and a total stellar mass created in the current episode of log($M[$M$_{\odot}]$)$ = 8.94 \pm 0.15$.  As has been found in other studies of LBGs \citep{shapley05}, the inclusion of IRAC photometry gives stellar population parameters similar to those derived with fits to optical/near-IR photometry alone \citep{ellingson96}, but with smaller uncertainties.  

With the addition of the rest-frame near-IR (IRAC) data, it is possible to look for a population of older, redder stars that would boost the IRAC photometry above what is expected from the young starburst alone.  \citet{bechtold98} use ISO mid-IR (6.7 \& 11.5 $\mu$m) photometry and conclude that no more than 10\% of the mass is from an old ($\sim2$ Gyr) population.  However, because the starburst in this system is $\sim$$10^7$ years old, the rest-frame near-IR is highly luminous due to red supergiants.  This significantly decreases the near-IR mass-to-light ratio relative to models at other ages (eg. $10^6$ or $10^8$ yrs).  Thus, a massive older stellar population can exist but negligibly affect the IRAC photometry.  We show an example in Figure \ref{fig:sedfit}, where we include an older (2 Gyr instantaneous burst) stellar population with more than two times the stellar mass of the younger stars ($\sim 2\times10^9$ M$_{\odot}$).  The best fit reddening of the older stellar population, $E(B-V)_{old}=0.27$, is somewhat less than the younger burst component, $E(B-V)_{young}=0.38$.  The rest-frame near-IR flux of the older population is still a factor of two lower than that of the less massive, younger population.  In this situation, the addition of the two components has nearly the same SED as the single component fit.   We conclude that $\sim 10^9$ M$_{\odot}$ has been produced by the current star formation episode thus far, but there can be twice that mass in older stars while still fitting the data reasonably well (reduced $\chi^2 = 2.5$ vs. 1.7 for single component fit), resulting in a maximum $M_{stars} \lesssim 3\times10^{9}$ M$_{\odot}$.  However, \citet{pettini02} show that the lack of N and Fe-peak elements in the ISM suggests that most of the stellar mass is less than 300 Myr old.  Therefore, it is likely that the mass from an old stellar population is significantly less than the maximum amount obtained from our two component fit.  (Note:  The stated masses are the current existing stellar mass and do not include the mass returned to the ISM via SNe.  For the old (2 Gyr) single burst model, the {\it original} total mass is equal to 1.37 times our stated values.  The correction to the young, $\sim$10 Myr models is negligible.)  

The velocity dispersions in atomic \citep[H$\alpha$,][]{teplitz00} and molecular \citep[CO,][]{baker04} gas both give virial masses of $\sim 10^{10}$ M$_{\odot}$.  \citet{baker04} derive a molecular gas mass of $M_{gas} = 6.6^{+5.8}_{-4.3}\times10^9$ M$_{\odot}$.  Therefore, our maximum stellar mass plus molecular gas mass does not exceed the maximum total mass.

\subsection{Far-Infrared Spectral Energy Distribution}
\label{ir_sed}

\begin{figure}
\epsscale{1.0}
\plotone{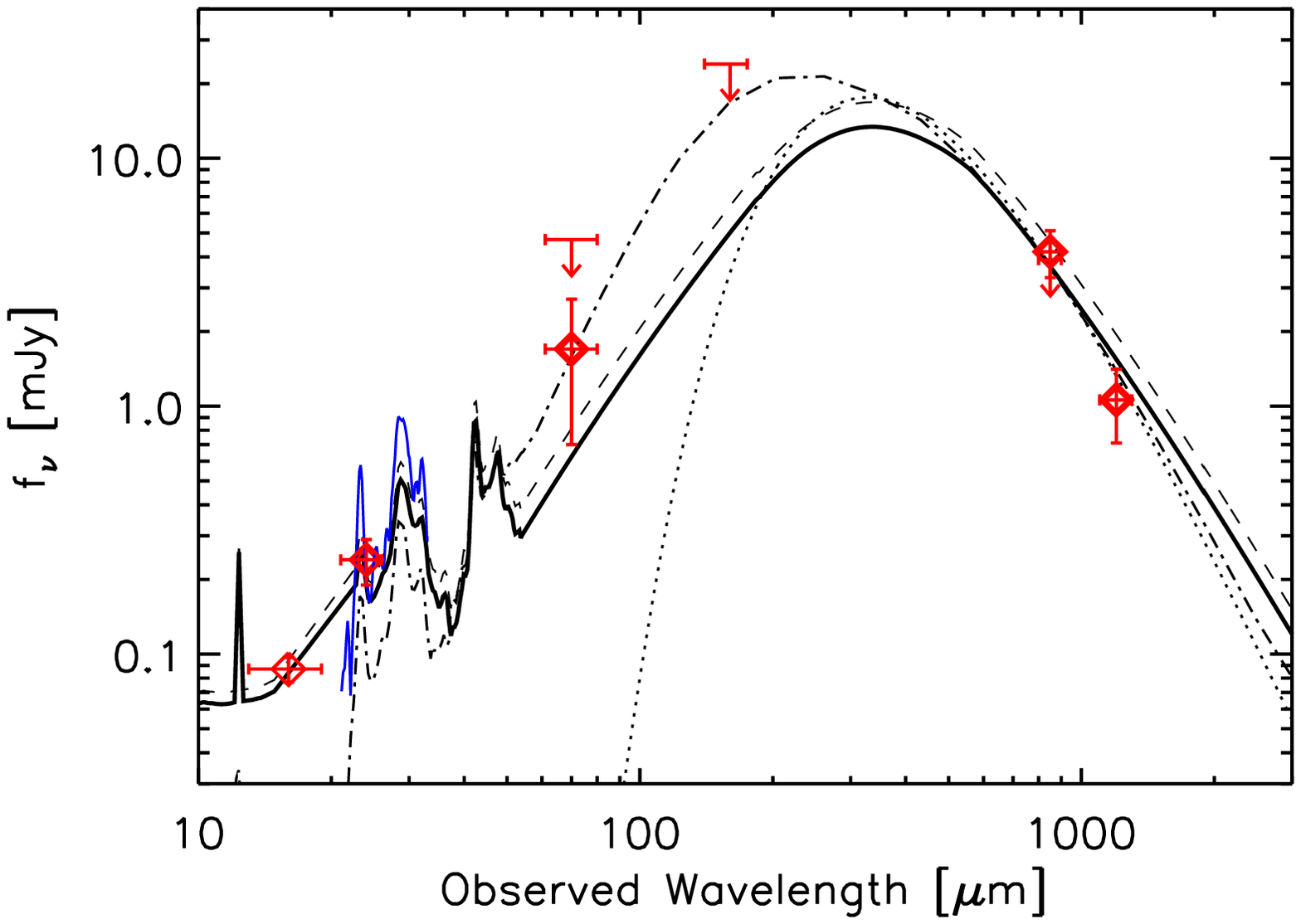}
\caption{The infrared SED of cB58.  The MIPS 16, 24 and 70 $\mu$m detections are plotted along with the conservative 70 and 160 $\mu$m limits.  The 850 $\mu$m detection and limit \citep[][respectively]{van_der_werf01,sawicki01} and the 1.2 mm detection \citep{baker01} are also plotted.  The best-fit SED from \citet{chary01} is plotted (solid line), giving a total $L_{IR} = 1.1\times 10^{11}$ L$_{\odot}$.  The dashed curve is the CE01 SED when picking a local template based on the $L_{24}$ alone, the dash-dotted curve is the best fit SED of \citet{dale02}, and the dotted curve is the IR SED assumed by \citet{baker01}.  The first-order Long-Low spectrum is shown in blue for reference.    \label{fig:ir_sed}}
\end{figure}

The MIPS photometry is plotted in Figure \ref{fig:ir_sed}.  We plot both the conservative 70 $\mu$m upper limit as well as the value derived from the three component fit.  The 850 $\mu$m \citep{van_der_werf01} and 1.2 mm \citep{baker01} detections are plotted as well.  Note that a $3\sigma$ upper limit of 3.9 mJy at 850 $\mu$m has also been reported \citep{sawicki01}, in disagreement with the nearly 5$\sigma$ reported detection $f_{850} = 4.2$ mJy by \citet{van_der_werf01}.  To estimate the IR luminosity, we fit the 105 infrared SEDs of \citet[][hereafter CE01]{chary01} to the detections at 24, 70, 850, and 1200 $\mu$m.  The best-fit model is shown as the solid curve in Figure \ref{fig:ir_sed}, The total infrared (8-1000 $\mu$m) luminosity is $L_{IR} = 1.1\pm0.1 \times 10^{11}$ L$_{\odot}$.  If we assume that the local templates of CE01 apply at high-z, and we simply use the template with the same $L_{24}$, we get an $L_{IR}=1.5\times 10^{11}$ L$_{\odot}$ (dashed line in Figure \ref{fig:ir_sed}).  This is actually the same template SED as in the complete SED fit, but with a higher normalization.  We also fit the SEDs of \citet[][hereafter DH02]{dale02} and get a best-fit $L_{IR} = 2.4 \times 10^{11}$ L$_{\odot}$, about two times higher than with the fit to the CE01 templates.  However, the DH02 fit is significantly worse than the best-fit CE01 models (reduced $\chi^2 = 4.8$ with DH02 vs. $1.6$ with CE01).  The DH02 models do not vary significantly in the predicted submm-to-24 $\mu$m flux ratio and every model underpredicts the $f_{24}$ given the 850 $\mu$m \& 1.2 mm fluxes.  We therefore adopt the best-fit CE01 model and IR-luminosity but estimate a significant uncertainty of $\sim 0.3$ dex.  The $L_{IR} < 2.4  \times 10^{11}$ L$_{\odot}$ is a reasonable upper limit since the best-fit DH02 template passes just under our $3\sigma$ 70 \& 160 $\mu$m upper limits.  

Finally we note that, although the CE01 models produce better fits than the DH02 models, {\it none} of these models has the large observed $f_{24}/f_{850}$ or $f_{24}/f_{1200}$ ratios observed in cB58.  The observed $f_{24}/f_{850}(cB58) = 0.057$ is larger than the maximum template values $f_{24}/f_{850}(CE01)_{max} = 0.055$ and $f_{24}/f_{850}(DH02)_{max} = 0.034$.  The observed $f_{24}/f_{1200}(cB58) = 0.23$ is larger than the maximum template values $f_{24}/f_{1200}(CE01)_{max} = 0.13$ and $f_{24}/f_{1200}(DH02)_{max} = 0.06$.  There are starbursts at low redshift that have higher $f_{24}/f_{850}$ ratios than cB58.  M82, for example, would have a ratio $f_{24}/f_{850} = 0.08$ at $z=2.73$, when using the compiled SED of \citet{galliano07}.  Therefore, the models simply do not display the full range of variability in the SEDs of local starbursts.

\subsection{Mid-Infrared Spectrum}
\begin{figure}
\epsscale{1.0}
\plotone{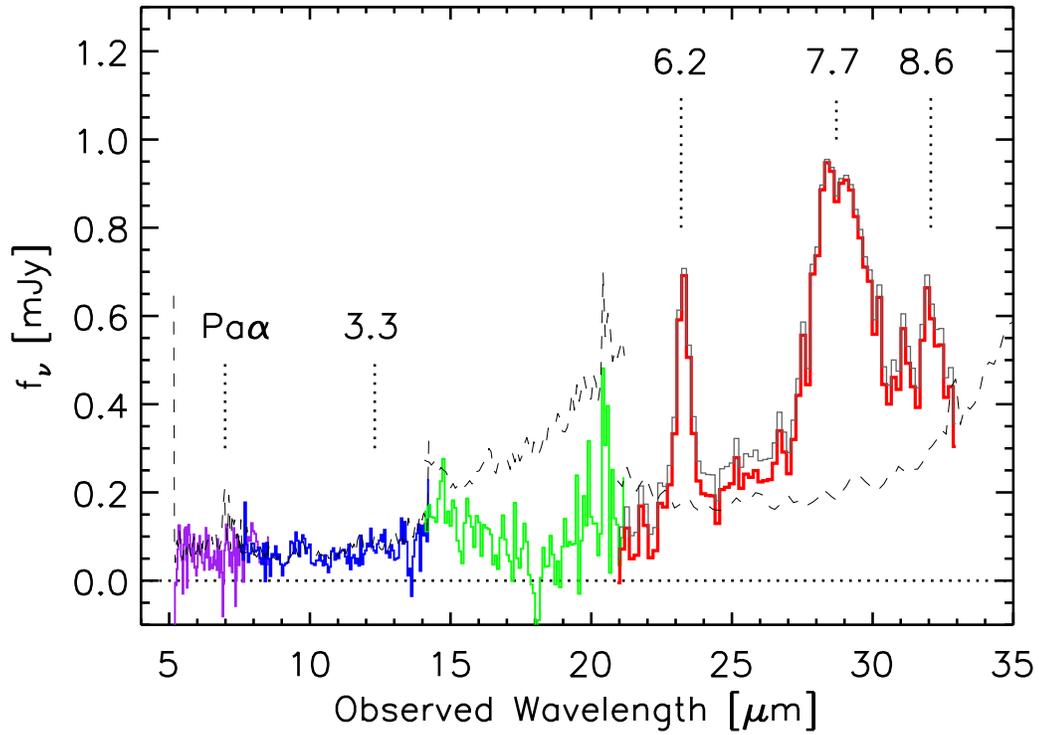}
\caption{The IRS spectrum of cB58.  Short-Low2, Short-Low1, Long-Low2, and Long-Low1 are plotted in purple, blue, green, and red solid lines, respectively (errors are plotted as dashed lines).  The wavelengths of Pa$\alpha$ and the primary PAH emission features are labeled. The original Long-Low1 spectrum before corrections for cD galaxy contamination and slit loss is plotted (thin solid grey line) to demonstrate how little these corrections affect the spectrum.  \label{fig:spectrum}}
\end{figure}

\begin{figure}
\epsscale{1.0}
\plotone{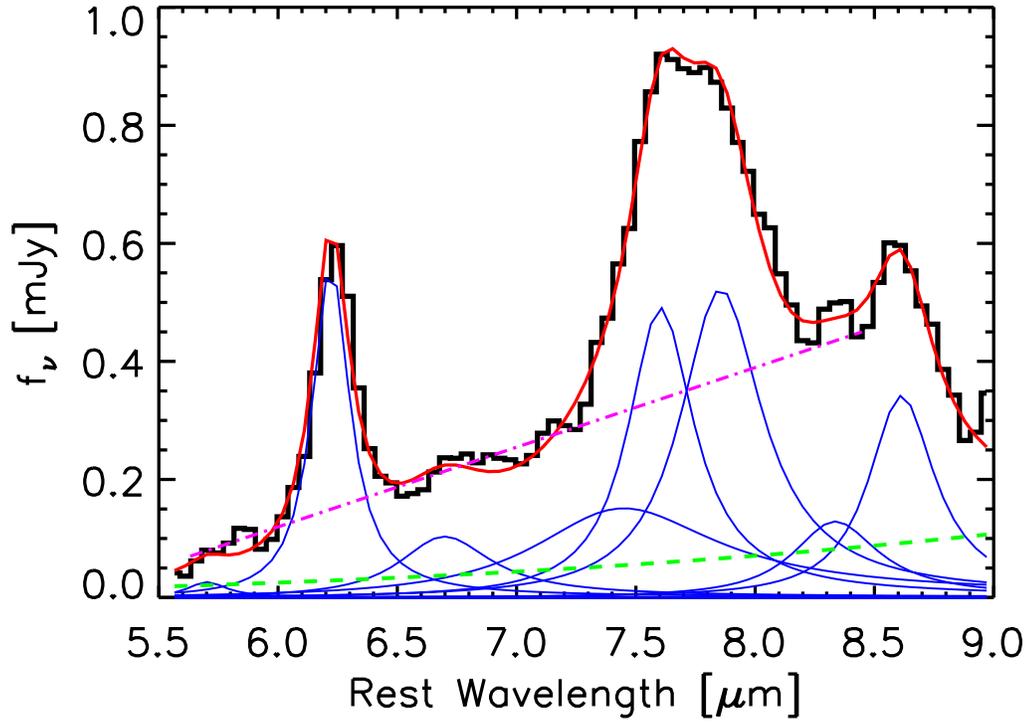}
\caption{Smoothed Long-Low 1st order spectrum of cB58 (black histogram) and the best-fit spectrum (red).  The best-fit individual PAH profiles (blue solid lines) and continuum (green dashed line) are also plotted.  The assumed continuum when comparing to the \citet{pope08} measurements is shown as a magenta dot-dashed line. \label{fig:pahfit}}
\end{figure}

The mid-infrared spectrum of cB58 is plotted in Figure \ref{fig:spectrum}.  As a reference, we have also plotted the Long-Low first order spectrum {\it before} the corrections for contamination and slit loss explained in Section \ref{obs_reduc}.  The corrections make only small differences to the overall spectrum between $20<\lambda<33$ $\mu$m.  Strong PAH emission is seen at 6.2, 7.7, and 8.6 $\mu$m.   The line strengths were measured by simultaneously fitting Drude profiles with centers and widths defined by \citet{draine07a}.  We also fit an underlying continuum and allow both the slope and amplitude to vary.  The resulting fit is plotted in Figure \ref{fig:pahfit}.  There is clearly large uncertainty in the amplitude of the minor PAHs, but the fluxes of the stronger PAHs (6.2, 7.7 complex, and 8.6) are well constrained.  The PAH fluxes are listed in Table \ref{tab:flux_pah}.  The continuum component is very low (the best-fit continuum comprises only $\sim 13$\% of the total flux measured between $5.7<\lambda_{rest}<9.0 \mu$m), and is not required for a good fit to the spectrum.  However, because we do not have data at $\lambda_{rest}> 9 \mu$m, we can not completely rule out a somewhat larger continuum component combined with silicate absorption at $\lambda_{rest} = 9.7 \mu$m.  

We do not detect the PAH emission feature at $\lambda = 3.3$ $\mu$m, putting a limit on the PAH ratio of $L_{6.2}/L_{3.3} > 3.9$.  This ratio depends strongly on the ionization state and grain size distribution of the PAHs \citep{li01, draine07a}. Assuming the grain size distribution and ionization fraction fit for Milky Way dust, the \citet{draine07a} models predict a luminosity ratio $L_{6.2}/L_{3.3}\sim3$, suggesting that we should see the line at $\sim 4\sigma$.    We note that \citet{draine07a} increased the strength of the 3.3 $\mu$m feature by a factor of 1.5-2.0 higher than the values in \citet{li01} to better fit theoretical calculations of the absorption cross section for various PAHs \citep{malloci07}.  

The best empirical dataset for flux ratios between the 6.2 and 3.3 $\mu$m features is that of \citet{imanishi06} and \citet{imanishi07}, where 6.2 $\mu$m is measured by {\it Spitzer} and 3.3 $\mu$m from the ground.  The sample contains nine ULIRGs that have no strong evidence (radio, optical, mid-IR) for an AGN {\it and} are compact enough (ie. no merging components) such that varying slit loss is a smaller issue.  These nine ULIRGs, have an average $<L_{6.2}/L_{3.3}> = 5.6$, with 7/9 having a $L_{6.2}/L_{3.3}$ larger than our $3\sigma$ limit.  Thus, if the PAH ratios in LBGs are similar to local ULIRGs it is not surprising that we did not detect the 3.3 $\mu$m PAH feature.  We note however that \citet{imanishi06} used a 0.9$''$ wide slit for the $L$-band spectroscopy, and the IRS Short-Low slit is 3.7$''$ wide.  The nine ULIRGs that we use here are 2-4$''$ across so there is a concern that the $L$-band spectroscopy is suffering more slit loss than the $Spitzer$ observations.  Therefore, if the spatial extent of the PAH emitting region is wider than the 0.9$''$ slit width, then the total $L_{6.2}/L_{3.3}$ ratios will be lower than the derived values, and may then disagree with the ratios measured in cB58.

\citet{moorwood86} measured 3.3 $\mu$m line fluxes for a number of starburst and Seyfert galaxies and compared them to the 8.6 $\mu$m fluxes available for seven of them.  These galaxies have lower luminosities than the ULIRG sample of \citet{imanishi07}.  The average luminosity ratio, $L_{8.6}/L_{3.3} = 2.1$, is higher than our $3\sigma$ lower limit of $L_{8.6}/L_{3.3}($cB58$)>1.9$ (as are 4 of the 7 individual ratios).  Thus, our limit is compatible with many of these lower luminosity galaxies as well as the ULIRGs.

Recently, \citet{magnelli08} find evidence for 3.3 $\mu$m PAH emission in 5 LIRGs at $0.638<z<0.839$.  In these five LIRGs they give a range of line-to-$L_{IR}$ ratios $L_{3.3}/L_{IR} < 1.29-3.5\times10^{-3}$.  The $3\sigma$ 3.3 $\mu$m limit for cB58 corresponds to $L_{3.3}/L_{IR} < 0.003$, so we would expect marginal (2-3$\sigma$) detections if the ratio were similar in cB58.

We also do not detect the Pa$\alpha$ line at $\lambda_{rest} = 1.87 \mu$m.  If we assume Case B recombination and scale from the H$\alpha$ flux with a reddening correction using the Calzetti reddening law and the gas reddening correction derived in \citet{baker01}, we find that the expected Pa$\alpha$ flux, $f_{Pa\alpha} = 2 \times 10^{-16}$ ergs s$^{-1}$ cm$^{-2}$.  This corresponds to about a 1$\sigma$ signal if all of the flux is contained within one pixel.  Thus, it is not surprising that we do not detect the Pa$\alpha$.

\section{Discussion}

\subsection{Star Formation Rate and Reddening}
\label{sfr}

There are now three independent indicators of the star formation rate of cB58: rest-frame UV continuum, HI recombination lines, and infrared luminosity.  When converting these measurements to star formation rates, we assume a Salpeter IMF with a 0.1-100 M$_{\odot}$ mass range.  \citet{pettini00}, adopting an LMC reddening law \citep{fitzpatrick86}, estimate a UV-derived $SFR(UV) = 93\pm11 $ M$_{\odot}$ yr$^{-1}$ (with a Salpter IMF between 0.1-100 $M_{\odot}$).  Continuous star formation for 100 Myr was assumed, but if a more realistic younger age of 9 Myr is assumed, the rate must be adjusted higher by 57\% \citep{kennicutt98} to $SFR(UV)=146\pm18$ M$_{\odot}$ yr$^{-1}$.  Our estimate based on the UV/optical/IRAC SED assuming a Calzetti reddening law is $SFR(UV)\sim 98^{+46}_{-31} $ M$_{\odot}$ yr$^{-1}$.  The H$\alpha$ line flux, corrected using a Calzetti reddening law and a color excess of the gas of $E(B-V)_{gas}=0.06$ derived from the \citet{teplitz00} Balmer decrement of H$\alpha$:H$\beta = 3.09$, gives $SFR(H\alpha) = 25\pm1$ M$_{\odot}$ yr$^{-1}$ using the conversion of \citet{kennicutt98}.  Finally, we convert our infrared luminosity to star formation rate using \citet{kennicutt98}, yielding $SFR(IR) = 19$ M$_{\odot}$ yr$^{-1}$ and a maximum of $SFR(IR)<41$ M$_{\odot}$ yr$^{-1}$ (assuming the best-fit DH02 model).  The derived SFR derivations are summarized in Table \ref{tab:sfr}.

The H$\alpha$ and IR-derived SFRs agree with each other, while the UV-derived SFR is higher by a factor of $3-5$.  Of course, the IR SFR assumes optically thick dust, such that the vast majority of the glaxy's bolometric luminosity is emitted in the IR.  For a UV-luminous galaxy such as cB58, we should consider the uncorrected UV star-formation rate (ie. using {\it observed} UV luminosity) in addition to the IR-derived SFR.  The uncorrected UV SFR, listed in Table \ref{tab:sfr} is $SFR(UV_{uncorr}) = 17 M_{\odot}$ $yr^{-1}$ when using the conversion of \citet{kennicutt98}.  Thus, even the addition of this component gives us a total SFR a factor of 2-4 times lower than the UV-derived SFRs.  Furthermore, cB58 is not significantly detected in a deep (50ks) Chandra image, also suggesting that the SFR must be lower than the UV-derived SFR (O. Almaini 2008, private communication).  It has been suggested previously that the discrepancy between the IR- and UV-derived SFRs is due to an overestimate of the reddening based on the UV continuum slope \citep{baker01,sawicki01}.  However, these previous $L_{IR}$ estimates were highly uncertain and could have been underestimated if the dust was significantly warmer than in local starbursts.  Our IR data at shorter wavelengths confirms that the dust temperature can not be much more than the 33 K assumed by \citet{baker01}.  In fact, our well constrained maximum $L_{IR}$ is still significantly lower than the UV SFR suggests it should be.  Therefore, the Calzetti reddening law is likely implying too high a UV extinction for cB58.  This is consistent with the analysis of \citet{reddy06}, who find that very young LBGs ($<$100 Myr) typically have UV slopes that overpredict their IR luminosities when using the relation derived by \citet{meurer99}, suggesting that a different extinction curve may be appropriate for young LBGs.

cB58 exhibits a damped Ly$\alpha$ profile, indicative of high HI column densities \citep{schaerer08}, and opaque low ionization metal absorption lines suggesting a covering fraction of neutral gas (and presumably dust) near unity \citep{pettini02,shapley03}.  Therefore, a reddening law for dust distributed in a uniform foreground screen may best describe the combined UV/IR properties of cB58.   The Calzetti reddening law is an empirical relation from many starburst galaxies with varying dust geometries, and is strongly weighted by the relatively unobscurred star-forming regions in these systems.  Thus, the Calzetti law may be more ``grey'' than is appropriate for galaxies with such large covering fractions of gas and dust, such as cB58.  When using line-of-sight extinction curves that are steeper in the UV \citep[SMC \& LMC,][ respectively]{prevot84, fitzpatrick86} to fit the stellar SED, we derive smaller star formation rates that are closer to the H$\alpha$, IR, and X-ray SFR derivations.  In fact, if we take out the 2175 \AA\ feature from the LMC extinction curve, we get a much better fit to the photometry than when using the Calzetti law (reduced $\chi^2 = 1.2$ and 1.7, respectively), as it reproduces the steep UV spectral slope (see inset in Figure \ref{fig:sedfit}).  The SFR derived when using LMC extinction (without the 2175 \AA\ feature) is 2.6 times lower than with a Calzetti law ($SFR_{LMC}(cB58) = 38$ M$_{\odot} yr^{-1}$, see Table \ref{tab:sfr} for a summary) and more in line with the other SFR determinations.  The better fit and realistic SFR when using the LMC curve, suggests that a steeper curve than the Calzetti law may be more appropriate for cB58.

This physical interpretation of a large covering fraction of neutral gas and dust in the foreground of cB58 may also help explain the discrepancies between UV- and IR-derived SFRs in other young $(t_{age}<100$ Myr) LBGs \citep{reddy06}.  The spectra of the youngest LBGs, like cB58, also exhibit higher equivalent width low-ionization interstellar absorption lines and redder UV spectral slopes than their older counterparts.  If the higher equivalent width absorption lines in these systems can be attributed to higher covering fractions, rather than higher velocity dispersions (the velocity structure is not resolved in the low-resolution spectra of unlensed LBGs), then a uniform foreground sheet dust geometry (and associated steeper extinction curves) may explain the relatively red UV spectral slopes of these young LBGs.

It is possible that the high frequency of Type II supernovae (SNe) in a young, intense, and compact starburst such as cB58 can significantly alter the composition of the dust from what is measured in more quiescent and older star-forming galaxies in the local universe (where extinction curves have been measured).  The extinction curves for Type II SNe dust \citep{maiolino04}, derived from dust compositions expected in Type II SNe \citep{todini01} have been invoked to explain the spectrum of a very high redshift BAL QSO \citep[z=6.2,][]{maiolino04} and the UV SED of a $z=6.29$ GRB afterglow \citep{stratta07}.  These systems are very young so their dust is therefore assumed to originate in Type II SNe.  The \citet{maiolino04} extinction curve is flat in the near-UV, but very steep at $\lambda < 1800$ \AA.  Using such an extinction curve, a red UV spectral slope between 1300 \AA\ $<\lambda<$ 1800 \AA\ implies less total dust absorption over the entire optical-UV spectrum than would be inferred with an SMC-like extinction curve or the Calzetti starburst reddening law.  Thus, an extinction curve of this nature would help explain the lower-than-expected $L_{IR}/L_{UV}$ because there is less total dust absorption, and would help explain the lower-than-expected $L_{H\alpha}/L_{UV}$ because the derived intrinsic far-UV flux (and thus $L_{H\alpha}$) would be lower.  However, our fit to the stellar SED with the \citet{maiolino04} extinction curve, though nicely fitting the red UV spectral slope, is a poor fit (reduced $\chi^2 = 3.6$) relative to the LMC and Calzetti fits as it does not fit the near-IR (rest-frame optical) photometry.  

A different IMF may also help explain the discrepancy in SFR indicators.  We have assumed a Salpeter IMF that extends up to 100 M$_{\odot}$.  If the IMF were deficient in the most massive O and B stars, the far-UV flux would be much lower than previously predicted, resulting in lower-than-expected $L_{H\alpha}$ and $L_{IR}$.  However, the rest-frame UV spectrum of cB58 is fit very well with stellar population models with a Salpeter IMF and an upper mass limit $M_{up}=100$ M$_{\odot}$ \citep{pettini00,pettini02}.  In particular, the observed P-Cygni profile of C IV requires the presence of stars with $M>50$ M$_{\odot}$ \citep{pettini00}.  Therefore, it is unlikely that a markedly different IMF is the cause of the discrepancy between UV- and IR/H$\alpha$-derived SFRs.  

\subsection{Infrared Comparison with Other Starbursts}

\subsubsection{Mid-IR Spectra Comparison}

\begin{figure}
\epsscale{1.0}
\plotone{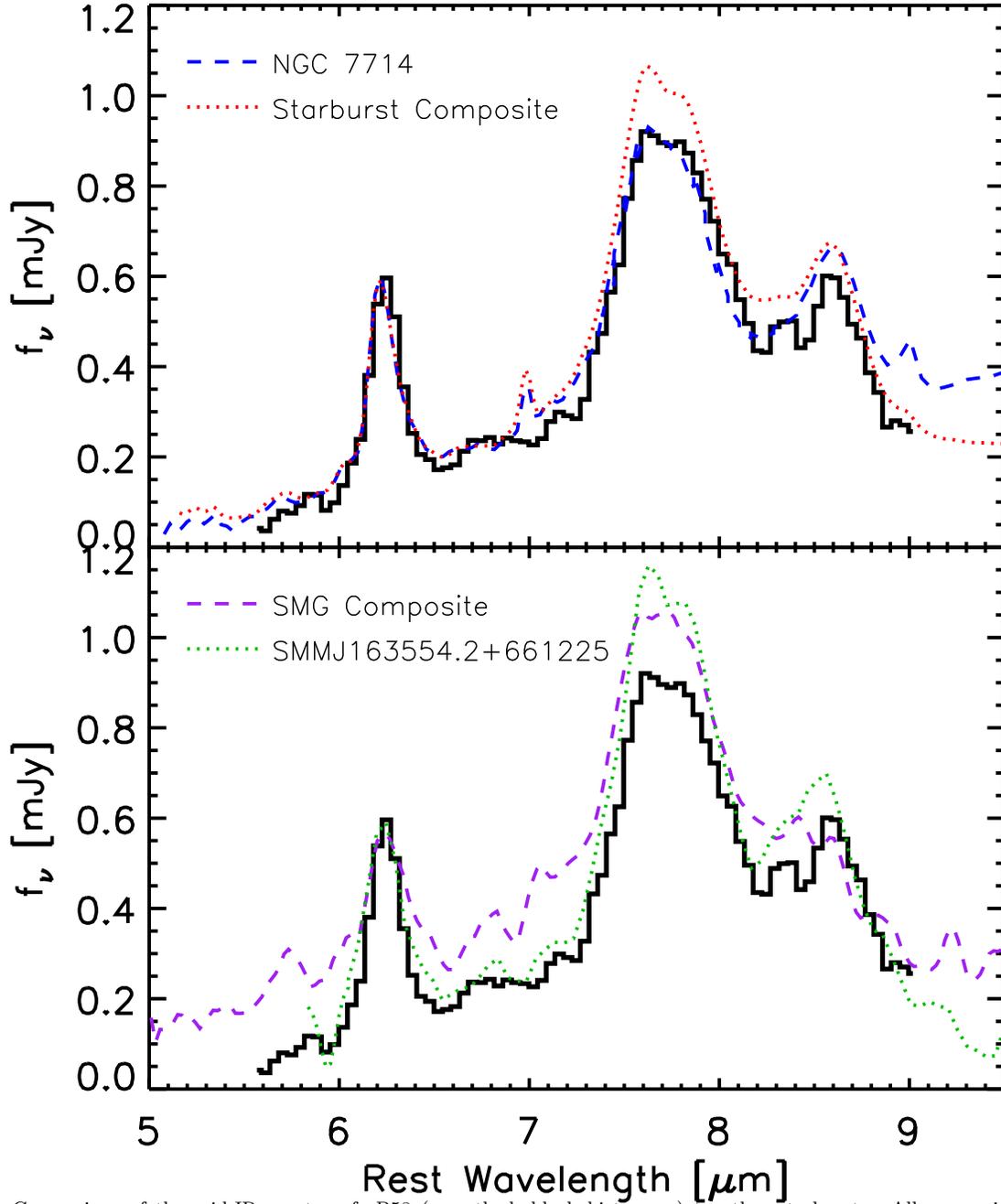}
\caption{Comparison of the mid-IR spectra of cB58 (smoothed, black histogram) to other starbursts.  All comparison spectra are normalized to have the same peak flux at 6.2 $\mu$m.  In the upper panel we compare to comparably bright local starbursts:  the spectrum of NGC 7714 \citep[blue dashed line,][]{brandl04} and the average starburst spectrum from \citet[red dotted line]{brandl06}.  In the lower panel we compare to high redshift starbursts: a composite spectrum of SMGs more than an order of magnitude more luminous than cB58 \citep[purple dashed line,][]{pope08}, and a lensed starburst with $4-6$ times the $L_{IR}$ of cB58 \citep[green dotted line,][]{rigby08}.  \label{fig:spec_comp}}
\end{figure}

In the upper panel of Figure \ref{fig:spec_comp}, we  plot the infrared spectrum of the core of NGC 7714 \citep[$L_{IR}=5.6\times 10^{10} $L$_{\odot}$,][]{brandl04} and a composite spectrum of 13 local starbursts \citep[$<$log($L_{IR}$)$> = 10.7$ L$_{\odot}$, ][]{brandl06}, scaled to match the flux of the 6.2 $\mu$m feature.  The IR luminosity of cB58 is 2-4 times higher than that of NGC 7714 and the starbursts of \citet{brandl06}.  The relative fluxes of the PAH features of cB58 agree very well with those seen in NGC 7714.  The $L_{7.7}/L_{6.2}$ ratio is a little higher in the starburst composite spectrum, but this makes little difference to the total 6-9 $\mu$m flux.  The PAH equivalent widths of cB58 are as large as in the local starbursts.  The exact continuum level is difficult to discern because of the large number of emission features and the limited wavelength coverage, but it is clear that the 6-9 $\mu$m luminosity is dominated by PAH emission.  

In the lower panel of Figure \ref{fig:spec_comp}, we compare the mid-IR spectrum of cB58 with high redshift, higher luminosity starbursts: a composite spectrum of 12 submm-selected galaxies by \citet{pope08} and a lensed, highly magnified submm source at $z=2.516$ with $L\sim8\times10^{11}$ L$_{\odot}$ \citep[SMM J163554.2+661225, ][]{rigby08}.  All templates are normalized to match the 6.2 $\mu$m emission strength.  The submm-selected galaxies are 1-2 orders of magnitude more luminous than cB58, and are dustier than UV-selected LBGs.  The mid-IR spectra look very similar both in their line strengths and the shape of the continuum, but few small differences can be seen.  First, as seen in comparing to the \citet{brandl06} composite of local starbursts, the $L_{7.7}/L_{6.2}$ is lower in cB58 than in the other spectra.  Second, the \citet{pope08} composite has a higher continuum to line ratio (especially at 6.2 $\mu$m) than that of cB58, likely due to hot dust emission from AGN, which may contribute up to 30\% of the mid-IR flux in SMGs \citep{pope08}.

\subsubsection{Mid- to Far-IR Comparison}

\begin{figure}
\epsscale{1.0}
\plotone{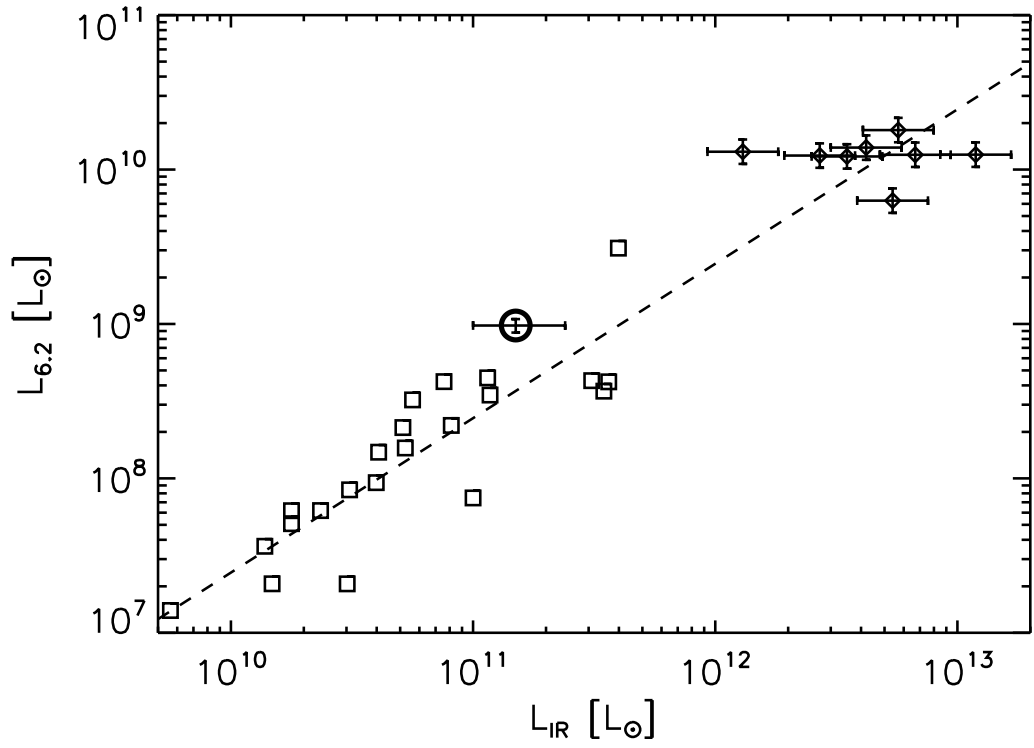}
\caption{The measured correlation between 6.2 $\mu$m and total infrared luminosities for local starbursts \citet[open squares,][]{brandl06} and high redshift SMGs \citet[open diamonds,][]{pope08}.  This figure is adapted from the top panel of Figure 12 in \citet{pope08}.  The best-fit relation to both populations is plotted as a dashed line.  cB58 (large open circle) lies above the relation such that it has more PAH luminosity relative to its infrared luminosity, but it lies within the scatter of the other starbursts.  \label{fig:pah_lir}}
\end{figure}

The PAH luminosities of pure starbursts have been shown to correlate with total infrared luminosity (and thus star formation rate) \citep{forster_schreiber04,peeters04,brandl06,pope08,rigby08}, though with considerable scatter.  This correlation has not yet been tested for high redshift galaxies with $L \lesssim 10^{12}$ L$_{\odot}$ such as cB58.  We compare the $L_{PAH}$ and $L_{IR}$ of cB58 to the relations fit by \citet{pope08} to local starbursts \citep{brandl06} and high-z submm-selected galaxies.  First, we re-measure the PAH luminosities in the same manner as \citet{pope08} by fixing the continuum levels at either side of the strong PAH features (as opposed to simultaneously fitting all PAH features as well as the continuum).  The assumed continuum is shown as the dot-dashed line in Figure \ref{fig:pahfit} and the corresponding luminosities are listed in Table \ref{tab:flux_pah}.  In Figure \ref{fig:pah_lir}, which shows the comparison between 6.2 $\mu$m luminosity and L$_{IR}$, we plot the cB58 position relative to the local starbursts and high-z SMGs.  The PAH luminosities lie on the high side of the measured relation such that the $L_{PAH}$ is high given the known $L_{IR}$, but the offset from the relation is within the scatter.  The 7.7 $\mu$m PAH luminosity is also higher than the fitted relation by the same factor.  The \citet{pope08} relations for both the 6.2 and 7.7 $\mu$m features both suggest an $L_{IR}\sim 3\times10^{11}$ L$_{\odot}$, about a factor of two higher than the best estimate of $L_{IR}$(cB58).

Another important relation is the correlation of the rest-frame 8 $\mu$m and total IR luminosities \citep{rigby08,bavouzet08,sajina08}.  This is particularly useful for broadband only studies at high redshift (eg. $Spitzer$ 24 $\mu$m) where individual PAH luminosities can not easily be measured.  From the IRS spectrum, we measure the rest-frame 8 $\mu$m luminosity (the luminosity measured through the IRAC 8 $\mu$m band at $z=0$) of cB58 to be $L_8(cB58) = 2.57\times10^{10}$ L$_{\odot}$.  Using the fitted relation of \citet{bavouzet08}, the $L_{8}(cB58)$ predicts an $L_{IR}(cB58)=1.7\times10^{11}$ L$_{\odot}$, in good agreement with our $L_{IR}$ estimate.

Finally, we compare the  $f_{24}/f_{850}$ flux ratio of cB58 to that of SMGs \citep{pope06} at similar redshifts.  The cB58 ratio, $f_{24}/f_{850} = 0.057$, is higher than six of the seven SMGs at similar redshifts ($2.3<z<3.1$), which range from $0.0031<f_{24}/f_{850}(SMGs)<0.062$.  The one SMG with a similar ratio has an X-ray hardness ratio indicative of an obscurred AGN that is likely contiributing significantly to the 24 $\mu$m flux.  \citet{pope06} suggest that additional extinction of the rest-frame mid-IR flux is required to reproduce the low $f_{24}/f_{850}$ flux ratios observed in SMGs.  Therefore, the higher $f_{24}/f_{850}$ of cB58 may simply be due to the lack of extinction in the rest-frame mid-IR relative to SMGs.  Additional extinction of SMGs in the mid-IR is expected, as these systems contain far more dust than LBGs and often exhibit severe extincion in the rest-frame UV/optical \citep[eg.][]{smail99}.   Alternatively, the discrepancy can be explained by more cold dust emission in SMGs relative to cB58.  This is perhaps not surprising as SMGs are selected on their cold dust emission.

\section{Conclusions}
We have obtained Spitzer photometry and mid-IR spectroscopy of the lensed LBG, MS1512-cB58.  This is the (intrinsically) least luminous galaxy to be studied in detail in the infrared at $z>1$.  Although cB58 is a ``typical'' LBG in terms of its UV and IR luminosities, the lack of a Balmer break in the SED indicates that it is much younger ($t_{age} \sim 10 Myr$) than a large majority of LBGs.  As seen in previous studies \citep{shapley05}, stellar population fits to the combined optical/near-IR/IRAC photometry give similar parameters to fits to the optical/near-IR photometry alone, but with smaller error bars.  The mid-IR photometry do not significantly constrain the mass of an older stellar population, as the rest-frame near IR flux from the current burst ($M_{burst} \sim 10^9$ M$_{\odot}$) can dwarf the flux from an older stellar population with twice the mass. 

The far-IR photometry is reasonably well fit by local starburst templates and gives an $L_{IR} = 1-2\times10^{11}$ L$_{\odot}$.  In addition, the PAH luminosities and rest-frame 8 $\mu$m luminosities agree with the $L_{PAH}$-to-$L_{IR}$ and $L_{8}$-to-$L_{IR}$ relations measured in local starbursts of comparable luminosity and high redshift, higher luminosity starbursts.  

The inferred SFR from the infrared luminosity is consistent with the SFR derived from the extinction corrected $H\alpha$ luminosity, but the UV-derived (extinction corrected) SFR is larger by a factor of 3-5.  This phenomenon has been noted by \citet{reddy06}, where LBGs with starburst ages less than 100 Myr consistently have lower $L_{IR}$ than the UV spectral slope and luminosity suggest.  With cB58, we can be confident that this discrepancy is {\it not} due to a poor determination of the infrared luminosity.  This suggests that the Calzetti obscuration law for starbursts may not be valid in very young, $<100$ Myr old, high redshift starbursts.  

The high HI column density and large covering fraction of neutral gas suggests a dust geometry approximated by a uniform foreground sheet.  This geometry would result in a steeper reddening law like the LMC or SMC curves and would explain the overstimate of the UV obscuration using the Calzetti reddening law.  In addition, because many young ($t_{age}<100$ Myr) LBGs also exhibit spectral properites indicative of large covering fractions of neutral gas (eg. high equivalent width low ionization metal absorption lines), a steeper reddening law may also be appropriate for explaining the relatively red UV spectral slopes observed in these systems as well.  

It is of course impossible to apply the relations measured in cB58 to the LBG population as a whole.  For example, there is significant scatter (beyond measurement errors) in the predicted opacities versus UV spectral slope in the starburst sample used to derive the Calzetti law \citep{calzetti94}.  Therefore, cB58 may simply lie to one end of the natural dispersion already observed in low redshift starbursts.  However, the fact that the phenomenon of the UV overpredicting $L_{IR}$ has been observed in a large sample of young LBGs \citep{reddy06} suggests that cB58 may be typical of these young sytems, and that a systematic bias in UV-derived SFR derivations may exist for the youngest LBGs.  SED fits to rest-frame UV and optical photometry of LBGs suggest that at least $\sim 30$\% have starburst ages less than 100 Myr \citep{shapley01,papovich01} and, at least for LBGs with $L_{UV}\sim L^*$, there is no correlation between age and observed UV luminosity \citep{shapley01}.  Therefore, if the SFR in all of these young LBGs is overestimated by a factor of $\sim 4-5$, then the global star formation rate density of all LBGs will be overestimated by a factor of $\sim 2$.  $Spitzer$ observations of recently discovered lensed LBGs \citep{smail07,coppin07,allam07} will allow us to better determine if these discrepancies persist in the rest of the young LBG population.  

\acknowledgments

We would like to thank Chuck Steidel and Omar Almaini for helpful discusions.  This work is based on observations made with the {\it Spitzer Space Telescope}, which is operated by the Jet Propulsion Laboratory, California Institute of Technology under a contract with NASA.  Support for this work was provided by NASA through an award issued by JPL/Caltech.

{\it Facilities:} \facility{Spitzer}.

\clearpage

\bibliographystyle{apj}
\bibliography{apj-jour,all_ref}

\begin{deluxetable}{lrr}
\tablecaption{$Spitzer$ Photometry of MS1512-cB58}
\tablehead{\colhead{Band} & \colhead{Flux Density} & \colhead{Error}\\
           \colhead{[$\mu$m]} & \colhead{[$\mu$Jy]} & \colhead{[$\mu$Jy]}}
\startdata
3.6		&	69.0 & 4.9 \\ 
4.5		&	77.5 & 5.5 \\ 
5.8		&	77.0 & 4.6 \\ 
8.0		&	53.7 & 3.8 \\ 
16      &   87 & 10 \\ 
24		&	240 	&	40.	\\
70		&   1.7 mJy 	&	1.0 mJy	\\
160\tablenotemark{a}		&	$<$24 mJy	&
\enddata
\tablenotetext{a}{3$\sigma$ confusion limit from \citet{dole04}}
\label{tab:photom}
\end{deluxetable}

\begin{deluxetable}{lcc}
\tablecaption{Star Formation Rate Estimates}
\tablehead{\colhead{SFR Indicator} & \colhead{SFR \tablenotemark{a}} & \colhead{$\chi^2/\nu$} \\
	       \colhead{} & \colhead{M$_{\odot}$ yr$^{-1}$} &  }
\startdata
UV	& \\
\hspace{4mm} SED Fit (This work)			 & & \\
	\hspace{8mm} Calzetti Law					 &  98$^{+46}_{-31}$  &  1.7  \\
	\hspace{8mm} LMC							 &  38$^{+24}_{-15}$  &  1.2  \\
\hspace{4mm} Pettini et al. 2000\tablenotemark{b} &	$146\pm18$\tablenotemark{c} &  \\
\hspace{4mm} Uncorrected $L_{1600}$ 		& $17\pm2$\tablenotemark{c} & \\ 
IR (This work)								 &  19-41\tablenotemark{c} & \\
H$\alpha$\tablenotemark{d}		 &  $25\pm1$\tablenotemark{c} & 
\enddata
\tablenotetext{a}{Corrected for 30x magnification}
\tablenotetext{b}{Corrected to use L$_{UV}$-to-SFR relation for age of 9 Myr, rather than 100 Myr}
\tablenotetext{c}{Errors are statistical and do not reflect uncertainties in conversion to SFRs}
\tablenotetext{d}{Extinction corrected assuming E(B-V)$_{gas} = 0.12$ \citep{teplitz00,baker01}}
\label{tab:sfr}
\end{deluxetable}

\begin{deluxetable}{lccc}
\tablecaption{Emission Line Fluxes of cB58}
\tablewidth{0pt}
\tablehead{\colhead{Wavelength} & \colhead{Flux \tablenotemark{a}} & \colhead{Luminosity \tablenotemark{a}} & \colhead{Rest Equivalent Width} \\
	        \colhead{[$\mu$m]} & \colhead{[$10^{-15}$ erg s$^{-1}$ cm$^{-2}$]} & \colhead{[$10^{44}$ erg s$^{-1}$]} & \colhead{[$\mu$m]}}
\startdata
1.87 (Pa$\alpha$) \tablenotemark{b}	&   	$<$0.60		& $<$0.37 &	\\ 
3.3	\tablenotemark{b}	&	$<$0.84	&   	$<$0.52	    & \\ 
6.2						&	3.3$\pm0.6$   &	2.0	(1.1)\tablenotemark{c} & 5.4	\\
7.7 (7.42+7.60+7.85)	&   10.0$\pm3.0$   &	5.8	(2.7)\tablenotemark{c} & 12	\\
8.6						&	1.6$\pm0.8$   &	1.0                        & 1.6
\enddata
\tablenotetext{a}{Not corrected for magnification}
\tablenotetext{b}{3$\sigma$ limit}
\tablenotetext{c}{Values in parentheses are the line luminosities derived when assuming the continuum shown as a dot-dashed line in Figure \ref{fig:pahfit}}
\label{tab:flux_pah}
\end{deluxetable}

\end{document}